\documentclass[seceq]{ptptex}

\usepackage{graphicx}
\usepackage{wrapft}
\usepackage{tabularx} 
\usepackage{multirow}
\usepackage{array}
\usepackage{varwidth}
\usepackage{booktabs}
\usepackage{tipa}
\usepackage{bm}
\def\TReg {\textsuperscript {\textregistered}}
\usepackage{textcomp}

\title{
Metric Deformation and Boundary Value Problems in 2D%
}

\author{
Subhasis \textsc{Panda}$^{1,}$\footnote{subhasis@cts.iitkgp.ernet.in}
Tapomoy  \textsc{Guha Sarkar}$^{2,}$\footnote{tapomoy@hri.res.in}
and S. Pratik \textsc{Khastgir}$^{3,1,}$\footnote{pratik@phy.iitkgp.ernet.in}
}

\inst{
$^1$Centre for Theoretical Studies, I.I.T. Kharagpur, 721302, India\\
$^2$Harish-Chandra Research Institute, Chhatnag Road, Jhusi, Allahabad, 211019, India\\
$^{3}$Department of Physics and Meteorology, I.I.T. Kharagpur, 721302, India

}

\abst{A new analytical formulation is prescribed to solve the
  Helmholtz equation in 2D with arbitrary boundary. A suitable
  diffeomorphism is used to annul the asymmetries in the boundary by
  mapping it into an equivalent circle. This results in a modification
  of the metric in the interior of the region and manifests itself in
  the appearance of new source terms in the original
  homogeneous equation. The modified equation is then solved
  perturbatively. At each order the general solution is written in a
  closed form irrespective of boundary conditions. This method allows
  one to retain the simple form of the boundary condition at the cost
  of complicating the original equation. When compared with numerical
  results the formulation is seen to work reasonably well even for
  boundaries with large deviations from a circle. The Fourier
  representation of the boundary ensures the convergence of the
  perturbation series.  }

\begin{document}

\maketitle

\section{Introduction}
The two dimensional Helmholtz equation appears in a wide range of
physical and engineering problems across diverse fields $-$ like the
study of vibration, acoustic and electromagnetic (EM) wave propagation
and quantum mechanics. In a large class of these problems one is
required to determine the eigenspectrum of the Helmholtz operator for
various boundary conditions and geometries. Canonical examples of the
Dirichlet boundary condition (DBC) are the vibration of membranes, the
propagation of the TM modes of EM waves within a waveguide and a
quantum particle confined in an infinite deep potential well. Perhaps
the prominent example of the Neumann boundary condition (NBC) is the
transmission of the TE modes of EM waves in a waveguide. Analytic
closed-form solution to the boundary value problems
\cite{b.1a,b.1b,b.1c} can however be obtained only for a restricted
class of boundaries. The problem for rectangular, circular, elliptical
and triangular boundaries are classical ones addressed by Poisson,
Clebsch, Mathieu and Lam{\'e} respectively \cite{b.1a}. Invoking the
geometry of the problem by a suitable choice of co-ordinates often
aids in finding the solutions (e.g. elliptic boundary where,
separation of variables leads to a solution in the form of Mathieu
functions). However, one quickly exhausts the list of such problems
where simplification by virtue of using a specific co-ordinate system
is possible. In most physical problems, one encounters boundaries
which are far removed from such idealization like the case of the
quantum dot. The dots are believed to be circular but in practice that
can hardly be guaranteed \cite{b.2,b.2a,b.2b,b.2c}. In such a scenario
it is natural to consider the confining region to be a supercircle
\cite{b.3}. Another important deviation from such idealization is the
design of waveguides with a shape, other than a rectangular or a
circular, that can be handy to purge the losses due to corners
\cite{b.w1}. Further, the problem gets analytically intractable for
arbitrary boundaries. The study of propagation of the
electromagnetic waves in open dielectric systems for an arbitrary
cross-section has been studied recently \cite{b.d1}.

The problem of solving Helmholtz equation for an arbitrary boundary
has mostly been tackled using numerical methods
\cite{b.4a,b.4b,b.4c,b.4d,b.4e,b.4f,b.4g,b.4h,b.4i,b.4j,b.4k,b.4l,b.4m,b.4n}. The
analytic approach towards this has mainly revolved around various
approximation methods. Of these, the perturbative techniques stand out
as being the most widely used
\cite{b.1a,b.1b,b.1c,b.4l,b.5a,b.5b,b.5c,b.5d,b.5e,b.5f}, where the
starting point is the Rayleigh's theorem : which states that the
gravest tone of a membrane whose boundary has slight departure from a
circle, is nearly the same as that of a mechanically similar membrane
in the form of a circle of the same mean radius or area. For slight
departure from circular boundary, one expands the wavefunction in
terms of a complete set of eigenfunctions (viz. Bessel functions) of
the unperturbed case (i.e. a circular boundary). Then the wavefunction
is made to satisfy the given condition on the arbitrary boundary and
which in turn extracts out the expansion coefficients of the required
wavefunction. The works \cite{b.1c,b.5e,b.6a,b.6b,b.6c} study
arbitrary domains in a general formalism using Fourier representation
of the boundary asymmetry treated as a perturbation around an
equivalent circle. In contrast ref. [24] studies analytic methods
where the arbitrary simply connected domain is mapped to a square by
conformal mapping and then the eigenvalues are approximated order by
order in the basis of a square boundary. In ref. [24] the zeros of
Bessel functions are approximated (i.e. energy eigenvalues associated
with a unit radius circle) from a square box. In our case we have
approximated the eigenvalues associated with a square box from an
equivalent circle. The ref. [24] has done a perturbation up to the
third order whereas in our case we have done up to first order except
the cases for $l = 0$ states where the second order corrections are
also included. It can be easily seen from the respective tables (Table
III in ref. [24] and Table I in our paper) that at the first order, both
the methods have comparable efficiency. Ref. [24] focuses its
attention mainly towards the eigenvalues in case of Dirichlet boundary
condition, whereas our formulation in a single stroke handles Neumann
condition as well. Moreover our paper gives at each order of
perturbation the correction to the wavefunction exactly. We have also
written down the exact expression for the $n^{\rm{th}}$ order correction to the eigenvalue in abstract
sense.

We have explored an alternative approach towards solving the
eigenvalue problem for the two dimensional Helmholtz operator in the
interior of a region bounded by an arbitrary closed curve. The general
problem is mapped into an equivalent problem where the boundary is a
regular closed curve (for which the Helmholtz equation is exactly
solvable) whereas the equation itself gets modified owing to the
deformation of the metric in the interior. The modified equation, we
see, can be written as the original Helmholtz equation with additional
terms arising from transformation in the metric. The extra pieces can
now be treated as a perturbation to the original Helmholtz
operator. The equation is thereby solved using the Schr\"{o}dinger
perturbation technique \cite{b.7}. The corrections to the
eigenfunctions are expressed in a closed form at each order of
perturbation irrespective of boundary condition. The eigenvalue
corrections are then obtained by imposing the appropriate boundary
condition. This is in contrast to the earlier methods where the
formulations are generally boundary condition dependent. In this
approach towards solving the equation, the boundary conditions are
specified on some known regular curve and maintain the same simple
form at each order of perturbation. This bypasses the issue of
imposing constraints on a boundary having a complicated geometry
\cite{b.6a,b.6b,b.6c}.

We expect this perturbative scheme to effectively solve the eigenvalue
problem for boundaries which reflect slight departure from known
regular curves. We have verified our method against the numerically
obtained solutions for a supercircle and an ellipse. In this analysis
we use Fourier representation of the boundary. This allows us to apply
the method to a general class of continuous or discontinuous
asymmetries. Section 2 describes the general formalism in abstract
sense. In section 3, we deal with the non degenerate case ($l = 0$)
and find solutions up to second order. Section 4 tackles the
degenerate case ($l \neq 0$) and obtains the first order correction to
energy and eigenfunction. In section 5, we apply the method for
supercircular and elliptical boundaries and compare our analytic
perturbative results with the numerically obtained ones. Finally, we
summarise our results noting the advantages of our method over the
other existing ones and conclude with a few comments.

\section{Formulation} 
The homogeneous Helmholtz equation on a 2 dimensional flat simply
connected surface $\cal S$ reads,
\begin {equation}
\left(g_{ij} \nabla^{i} \nabla^{j} + k^{2}\right)\psi \equiv \left(\nabla^{2} + E\right) \psi=0,   \label{eq:1}
\end {equation}
where $g_{ij}$ is the flat metric and $\nabla$ represents a covariant
derivative. We look for solutions in the interior of the bounded
region with the Dirichlet condition $\psi=0$ or the Neumann condition
$\frac{\partial \psi}{\partial n} = 0 $ on $\partial \cal S$, where
$\frac{\partial \psi}{\partial n}$ denotes the derivative along the
normal direction to $\partial \cal S$. The parameter $k^{2}$ may be
identified with $E$, the energy of a quantum particle confined in the
region having a boundary $\partial \cal S$.

It is convenient to work in polar coordinate system ($r, \theta$),
where any closed curve satisfies the periodicity condition
$r(\theta)=r(\theta + 2\pi)$. We consider a general arbitrary boundary
of the form $r = r(\theta)$. In this analysis we assume that the
arbitrary boundary can be expressed as a perturbation around an
effective circle (the analysis can in principle work for a deformation
around any simple curve for which the Helmholtz equation is exactly
solvable). We introduce new coordinates $(R, \alpha)$ with the
transformation $(r, \theta) \rightarrow (R, \alpha)$ given by
\begin{equation}
\label{eq:3}
\begin{split}
r =~ & R + \epsilon f(R,\alpha) ~;\\
\theta =~ & \alpha ~,
\end{split}
\end{equation}
where $\epsilon$ is a deformation parameter. This defines a
diffeomorphism for the entire class of well behaved functions
$f(R,\alpha)$. A suitable choice of $f(R,\alpha)$, shall transform our
arbitrary boundary into a circle of average radius, say $R_{0} \,
(=\frac{1}{2 \pi}\int \limits^{2 \pi}_{0} r(\theta) \,\,
\text{d}\theta)$ in the $R-\alpha$ plane. The deformation of the
arbitrary boundary to a circle changes the components of the
underlying metric $g_{ij}(r, \theta)$ in the interior ($g_{ij}(r,
\theta) \rightarrow \tilde{g}_{ij}(R, \alpha)$). Henceforth, we use
the notation $$ \phi^{(i,j)} \equiv \frac{\partial^{i+j}\phi}{\partial
  R^{i} \partial \alpha^{j}}~,$$ where $\phi$ is a function of $R$ and
$\alpha$. The dependence on the arguments $(R, \alpha)$ are not shown
explicitly for brevity.  The flat background metric in the ($r,
\theta$) system is given by $g_{ij}= {\rm diag} (1,r^{2})$. Under the
coordinate transformations ({\ref{eq:3}}) this takes the form
\begin{equation}
\tilde{g}_{ij} = \left[\begin{array}{ll}
 \quad\left(1 + \epsilon  f^{(1,0)} \right)^2  \qquad\quad\,\,\,~ \epsilon  f^{(0,1)}
 \left(1 + \epsilon  f^{(1,0)}\right) \\
 \epsilon f^{(0,1)} \left(1 + \epsilon  f^{(1,0)}\right)  \qquad (R
 +\epsilon f)^{2} + \epsilon^2 f^{{(0,1)}^{2}} \end{array}
\right],\nonumber 
\end{equation}
We note that except for $\Gamma^{\alpha}_{{\phantom \alpha} R R }$ all
the components of connection $\Gamma$ are non-vanishing. The
diffeomorphism ({\ref{eq:3}}) does not induce any
spurious curvature in the manifold (i.e. Riemann tensor,
$R^{i}_{{\phantom i} jkl} = 0~~\forall~ i, j, k, l$).

The Eq. ({\ref{eq:1}}), where $\psi = \psi(r, \theta)$, transforms
under the map $(r, \theta) \rightarrow (R, \alpha)$ to
\begin{flalign}
\label{eq:6}
&E ~\psi + \frac{\psi ^{(0,2)}}{(R+\epsilon f)^2}+\frac{\left[(R+\epsilon f)^2
      - \epsilon f^{(0,2)} \left(R + \epsilon f\right) + 2 \epsilon ^2
      f^{{(0,1)}^2} \right] \psi ^{(1,0)}}{(R+\epsilon f)^3
    \left(\epsilon f^{(1,0)}+1\right)} \nonumber\\ 
&+\frac{2 \epsilon^2 f^{(0,1)} \left[(R+\epsilon f) f^{(1,1)}-f^{(0,1)} \left(\epsilon
      f^{(1,0)}+1\right)\right]\psi ^{(1,0)} }{(R+\epsilon f)^3
    \left(\epsilon f^{(1,0)}+1\right)^2}-\frac{2 \epsilon f^{(0,1)} \psi
    ^{(1,1)} }{(R+\epsilon f)^2 \left(\epsilon
    f^{(1,0)}+1\right)} \nonumber\\ 
&-\frac{\epsilon f^{(2,0)} \left[(R + \epsilon f)^{2} + \epsilon^2 f^{{(0,1)}^2}\right] \psi
    ^{(1,0)} }{(R+\epsilon f)^2 \left(\epsilon
    f^{(1,0)}+1\right)^3}+\frac{\left[(R +
      \epsilon f)^{2} + \epsilon^2 f^{{(0,1)}^2}\right] \psi
    ^{(2,0)}}{(R+\epsilon f)^2 \left(\epsilon f^{(1,0)}+1\right)^2} = 0\,. 
\end{flalign}
The analysis can proceed from here for a specific form of the function
$f(R,\alpha)$. We choose $f(R,\alpha)= R g(\alpha)$, where $g(\alpha)$
can be expanded without a loss of generality in a Fourier series. We
further impose $g(\alpha) = g(-\alpha)$ for simplicity whereby only
the cosine terms are retained
\begin{equation}
g(\alpha) = \sum_{n=1}^{\infty} C_{n} \cos n\alpha \label{eq:19}.
\end{equation}
The constant part $C_{0}$ can always be absorbed in $R$ defined in
({\ref{eq:3}}). With this choice of $f(R,\alpha)$, Eq. ({\ref{eq:6}})
simplifies to
\begin{equation}
\sum_{n = 0}^{\infty} \epsilon^{n} {\cal L}_{n}\psi  +   E \psi = 0, \label{eq:7}
\end{equation}
where the operator ${\cal L}_{n}$ is given by 
\begin{flalign}
&{\cal L}_{n}\psi = (-1)^{n}\frac{(n+1)}{6 R^{2}} g^{n-2}\left[ 3 n R
  g \left\{ g^{(0,2)} \psi^{(1,0)} + 2 g^{(0,1)}
  \psi^{(1,1)}\right\}  \right.  \\& \left.\quad+ n (n -1) R \left(g^{(0,1)}\right)^2 \left\{ 2
  \psi^{(1,0)} + R \psi^{(2,0)} \right\}
  +6 g^{2} \left\{\psi^{(0,2)} + R \psi^{(1,0)} + R^{2}
  \psi^{(2,0)} \right\} \right].\nonumber
\end{flalign}

We shall adopt the method of stationary perturbation theory \cite{b.7}
to solve for $\psi$ and $E$.  Thereby, treating $\epsilon$ as a
perturbation parameter we expand the eigenfunction $\psi$
corresponding to the eigenvalue $E$ as
\begin{subequations}
\begin{align}
\psi &= \psi^{(0)} + \epsilon \psi^{(1)} + \epsilon^{2} \psi^{(2)}
+\cdots ;  \\
E &= E^{(0)} + \epsilon E^{(1)} + \epsilon^{2} E^{(2)} +\cdots, 
\end{align}\label{eq:12}
\end{subequations}
with superscripts denoting the order of perturbation. We assume that
the perturbative scheme converges and $(\psi, E)$ can be calculated
order by order up to any arbitrarily desired precision.  We note that
the parameter $\epsilon$ is arbitrarily invoked to track different
orders and could be absorbed in the Fourier coefficients $C_{n}$.

Plugging ({\ref{eq:12}}) in (\ref{eq:7}), and collecting the
coefficients for different powers of $\epsilon$ yields
\begin{subequations}
\begin{align}
{\mathcal{O}}(\epsilon^{0})&: &&({\cal L}_{0} + E^{(0)}) \psi^{(0)} = 0~,\label{eq:13} \\  
{\mathcal{O}}(\epsilon^{1})&:  &&({\cal L}_{0}+ E^{(0)}) \psi^{(1)} + ({\cal L}_{1}
 + E^{(1)}) \psi^{(0)} = 0~, \label{eq:14}\\ 
{\mathcal {O}}(\epsilon^{2})&: &&({\cal L}_{0} +E^{(0)})  \psi^{(2)} +
({\cal L}_{1}+ E^{(1)}) \psi^{(1)}+({\cal L}_{2} + E^{(2)}) \psi^{(0)}  =
0~, \label{eq:15} \\
\vdots \nonumber\\
{\mathcal{O}}(\epsilon^{m})&:  &&\sum_{n = 0}^{m}\left({\cal L}_{n} + E^{(n)}\right)
\psi^{(m-n)} = 0~.\label{eq:15a}
\end{align}
\label{eq:15tot}
\end{subequations}
The change in the metric components induced by the smooth deformation
({\ref{eq:3}}) amounts to a gauge transformation and
generates source terms to the unperturbed homogeneous Helmholtz
equation at each order in $\epsilon$. At the $i^{\text{th}}$ order we have
the terms ${\cal L}_{1} \psi^{(0)},\, {\cal L}_{1} \psi^{(1)},\,
\cdots,\, {\cal L}_{i} \psi^{(j(<i))}$, which have no physical
origin and are merely artifacts of the chosen gauge. Maintaining the
simplicity of the boundary conditions is hence achieved at the
cost of new terms appearing in the original equation.

The unperturbed energy $E^{(0)}$ and corrections $E^{(1)}$, $E^{(2)}$
are given by
\begin{subequations}
\begin{align}
E^{(0)} = & -\langle \psi^{(0)}|{\cal L}_{0}|\psi^{(0)} \rangle ; \label{eq:16a} \\
E^{(1)} = & -\langle \psi^{(0)}|{\cal L}_{1}|\psi^{(0)}\rangle; \label{eq:16b}   \\
E^{(2)} = & -\langle \psi^{(0)}|{\cal L}_{1} + E^{(1)} |\psi^{(1)} \rangle -
  \langle \psi^{(0)}|{\cal L}_{2}|\psi^{(0)} \rangle. \label{eq:16} \\
\vdots \nonumber\\
E^{(m)} = & -\Big{\langle}
\psi^{(0)}\Bigl\lvert\sum_{n=1}^{m-1}\left({\cal L}_{n} +
E^{(n)}\right)\Bigr\rvert \psi^{(m-n)} \Big{\rangle} -
  \Big{\langle} \psi^{(0)} \Bigl\lvert {\cal L}_{m} \Bigr\rvert \psi^{(0)} \Big{\rangle} . \label{eq:16c}
\end{align} 
\label{eq:16tot}
\end{subequations}
A unique feature of our method is that both the boundary conditions
maintain their simple forms separately for every order in
perturbation. Thus, we have for the $i^{\text{th}}$ order wavefunction
the DBC and the NBC respectively,
\begin{subequations}
\begin{align}
&~\psi^{(i)}(R_{0},\alpha) = 0\,, \qquad\qquad\qquad\qquad\quad\qquad ~\mbox{(DBC)}\,;\label{eq:dbc}\\ 
&\left(\frac{\partial\psi^{(i)} }{\partial R} -\frac{g^{(0,1)}}{R}\frac{\partial\psi^{(i-1)} }{\partial \alpha}\right)\Bigg\rvert_{(R_{0},\alpha)} = 0\,, \quad\quad~\mbox{(NBC)}\,,\label{eq:nbc}
\end{align}\label{eq:bctot}
\end{subequations}
where $i \in \mathbb{N}$.
The general solution of the Eq. ({\ref{eq:13}}) is
\begin{align}
\psi_{l,j}^{(0)}  & =  N_{0,j} J_{0}(\rho)\,,  \qquad\qquad \qquad \qquad ~(l = 0)\,; \nonumber \\
         & =  N_{l,j} J_{l}(\rho)\left\{\begin{array}{c} \cos (l\alpha)
\\ \sin  (l\alpha) \end{array} \right\}, \qquad \quad  (l \neq 0)\,, 
         \label{eq:21}
\end{align}
where $J_{l}$ is the $l^{\text{th}}$ order Bessel function with the
argument $\rho = \sqrt{E_{l,j}^{(0)}} R$, where $E_{l,j}^{(0)}$ are
the energies of the unperturbed Helmholtz equation. $N_{l,j}$ is a
suitable normalisation constant with $l \in \mathbb{N}$, $j \in
\mathbb{N}_{>0}$. It is to be noted that the normalisation constant
will be different for the different boundary conditions. Henceforth,
we will discuss both the cases, viz. the Dirichlet and the Neumann
boundary condition parallely. The energy $E_{l,j}^{(0)}$ is dictated
by the $ j^{\text{th}}$ zero of $J_{l}$, denoted by $\rho_{_{l,j}}$,
and the $ j^{\text{th}}$ zero of $J^{\prime}_{l}$, denoted by
$\rho^{\prime}_{_{l,j}}$, for DBC and NBC respectively. Using
Eq. ({\ref{eq:dbc}}) and ({\ref{eq:nbc}}) for $i = 0$, we have
\begin{align}
 E_{l,j}^{(0)} =~&\rho^{2}_{_{l,j}}/R_{0}^{2} \,,~\qquad \qquad \mathrm{(DBC)} ~;\label{eq:endbc}\\
             =~&\rho^{\prime^{2}}_{_{l,j}}/R_{0}^{2}\,,\qquad \qquad ~\mathrm{(NBC)}~,\label{eq:ennbc}
\end{align}
where all the levels with non-zero $l$ are doubly degenerate.

In this formulation the energy corrections can be obtained in two
ways. Firstly, $E^{(i)}$ can be estimated from the knowledge of
$\psi^{(m)}(\forall~ m < i)$ using
Eqs. (\ref{eq:16tot}). Alternatively it can be extracted by imposing
the boundary condition on $\psi^{(i)}$ given by Eqs. (\ref{eq:bctot}),
which in addition yields the coefficients of Bessel functions (in
$\psi^{(i)}$). The method can in principle be used to calculate
corrections at all orders of perturbation. We next calculate the
energy corrections for both the boundary conditions for the following
two cases.
\section{Case I: Non-degenerate states ($l=0$)} 
The first order correction to the eigenfunction is obtained solving
the Eq. ({\ref{eq:14}}). Thus, we have
\begin{align}
\psi_{0,j}^{(1)} =&\, a_{0} J_{0}(\rho) -\frac{\rho E_{0,j}^{(1)}}{2
    E_{0,j}^{(0)}}N_{0,j}J_{1}(\rho) + \sum_{p=1}^{\infty}{\Big\{} a_{p} J_{p}(\rho)-\rho N_{0,j} C_{p}  J_{1}(\rho){\Big\}} \cos (p\alpha),   \label{eq:22} 
\end{align}
where $E_{0,j}^{(1)}$, $a_{0}$ and $a_{p}$ are constants to be fixed
by the boundary conditions. The terms contain $N_{0,j}J_{1}(\rho)$ make the particular integral of the Eq. ({\ref{eq:14}}). The first order energy corrections are
obtained by imposing the respective boundary condition, given by
Eq. ({\ref{eq:dbc}}) or by substituting $\psi_{0,j}^{(1)}$ and
$\psi_{0,j}^{(0)}$ into the Eq. ({\ref{eq:nbc}}), (for $i = 1$). This
yields
\begin{align*}
&E_{0,j}^{(1)} = 0 ~~~\mbox{(for both the cases)}; \\
&a_{p} = \rho_{_{0,j}} N_{0,j}
C_{p}J_{1}(\rho_{_{0,j}})/J_{p}(\rho_{_{0,j}})\,, ~ (p \neq 0)~\mbox{(DBC)};\\
&a_{p} = \rho^{\prime}_{_{0,j}} N_{0,j} C_{p}
J_{0}(\rho^{\prime}_{_{0,j}})/J^{\prime}_{p}(\rho^{\prime}_{_{0,j}})\,,
~ (p \neq 0)~\mbox{(NBC)}.
\end{align*}
The vanishing of the first order correction, $E_{0,j}^{(1)}$, is
verified using Eq. ({\ref{eq:16b}}). The remaining constant $a_{0}$ of
Eq. ({\ref{eq:22}}) is zero for both the boundary conditions by virtue
of orthogonality of $\psi_{0,j}^{(0)}$ and $\psi_{0,j}^{(1)}$. These
results are consistent with the results obtained in
\cite{b.6a,b.6b,b.6c} by other methods.

The first non-vanishing energy correction occurs at the second
order. The correction $E_{0,j}^{(2)}$ is obtained by substituting
$\psi_{0,j}^{(0)}$ and $\psi_{0,j}^{(1)}$ in Eq.  ({\ref{eq:16}}) and
have
\begin{align}
E_{0,j}^{(2)} = E_{0,j}^{(0)}\sum_{n=1}^{\infty} \xi_{n,j}  C^{2}_{n}  \,; ~~
\xi_{n,j}  = \frac{1}{2} +  \frac{\rho_{_{0,j}} J^{\prime}_{n}(\rho_{_{0,j}})}{J_{n}(\rho_{_{0,j}})},\label{eq:27}
\end{align}
and
\begin{align}
E_{0,j}^{(2)} = -\,E_{0,j}^{(0)} \sum_{n=1}^{\infty} \lambda_{n,j}  C^{2}_{n}  \,; ~
\lambda_{n,j}  = \frac{1}{2} +  \frac{\rho^{\prime}_{_{0,j}} J_{n}(\rho^{\prime}_{_{0,j}})}{J^{\prime}_{n}(\rho^{\prime}_{_{0,j}})},\label{eq:27nb}
\end{align}
for the DBC and the NBC respectively.  We may as well solve
({\ref{eq:15}}) to obtain
\begin{alignat}{2}
&\psi_{0,j}^{(2)} =  b_{0} J_{0}(\rho) - \frac{\rho E_{0,j}^{(2)}}{2
    E_{0,j}^{(0)}}N_{0,j}J_{1}(\rho)+ \sum_{n=1}^{\infty}C_{n}{\cal J}_{n,j}(\rho)\nonumber\\ 
&+\sum_{p=1}^{\infty}{\Big\{} b_{p} J_{p}(\rho)-\rho a_{0} C_{p}
        J_{1}(\rho) +\sum_{n=1}^{\infty}\left(C_{n+p} + C_{|n-p|}
        \right){\cal J}_{n,j}(\rho) {\Big\}}  \cos (p\alpha)\,,  
\label{eq:24}
\end{alignat}
where 
\begin{equation}
{\cal J}_{n,j}(\rho)=\frac{\rho}{2}{\Big\{} a_{n}
J^{\prime}_{n}(\rho)-\frac{\rho}{2} N_{0,j} C_{n}
J^{\prime}_{1}(\rho){\Big\}}.\nonumber
\end{equation}
Boundary conditions, Eqs. (\ref{eq:bctot}) (for $i = 2$), extract
$E_{0,j}^{(2)}$ as given in Eq. (\ref{eq:27}) or Eq. (\ref{eq:27nb})
and in addition the coefficients $b_{p}$ are respectively given by
\begin{align*}
b_{p} = & ~\frac{\rho_{_{0,j}} J_{1}(\rho_{_{0,j}})}{J_{p}(\rho_{_{0,j}})}\left\{a_{0}C_{p}-\frac{N_{0,j}}{2}\sum_{n=1}^{\infty}C_{n}(C_{n+p}+ C_{|n-p|})\xi_{n,j}\right\},~~~\mbox{(DBC)};~\\
 = & ~\frac{\rho^{\prime}_{_{0,j}} J_{0}(\rho^{\prime}_{_{0,j}})}{J^{\prime}_{p}(\rho^{\prime}_{_{0,j}})}\left\{a_{0}C_{p}+\frac{N_{0,j}}{2\rho^{\prime}_{_{0,j}}}\sum_{n=1}^{\infty}
np\,C_{n}(C_{n+p} - C_{|n-p|}) \right. \\ 
&\left. \qquad \qquad \qquad \qquad +\frac{N_{0,j}}{2}\sum_{n=1}^{\infty}C_{n}(C_{n+p}
+C_{|n-p|})\lambda_{n,j}\right\},~~~~ \mbox{(NBC)}.
\end{align*}
The remaining constant $b_{0}$ can be fixed by
normalising the corrected wavefunction.
\section{Case II: Degenerate states ($l \neq 0$)}
\begin{figure}
    \centerline{\scalebox{0.4}{\includegraphics{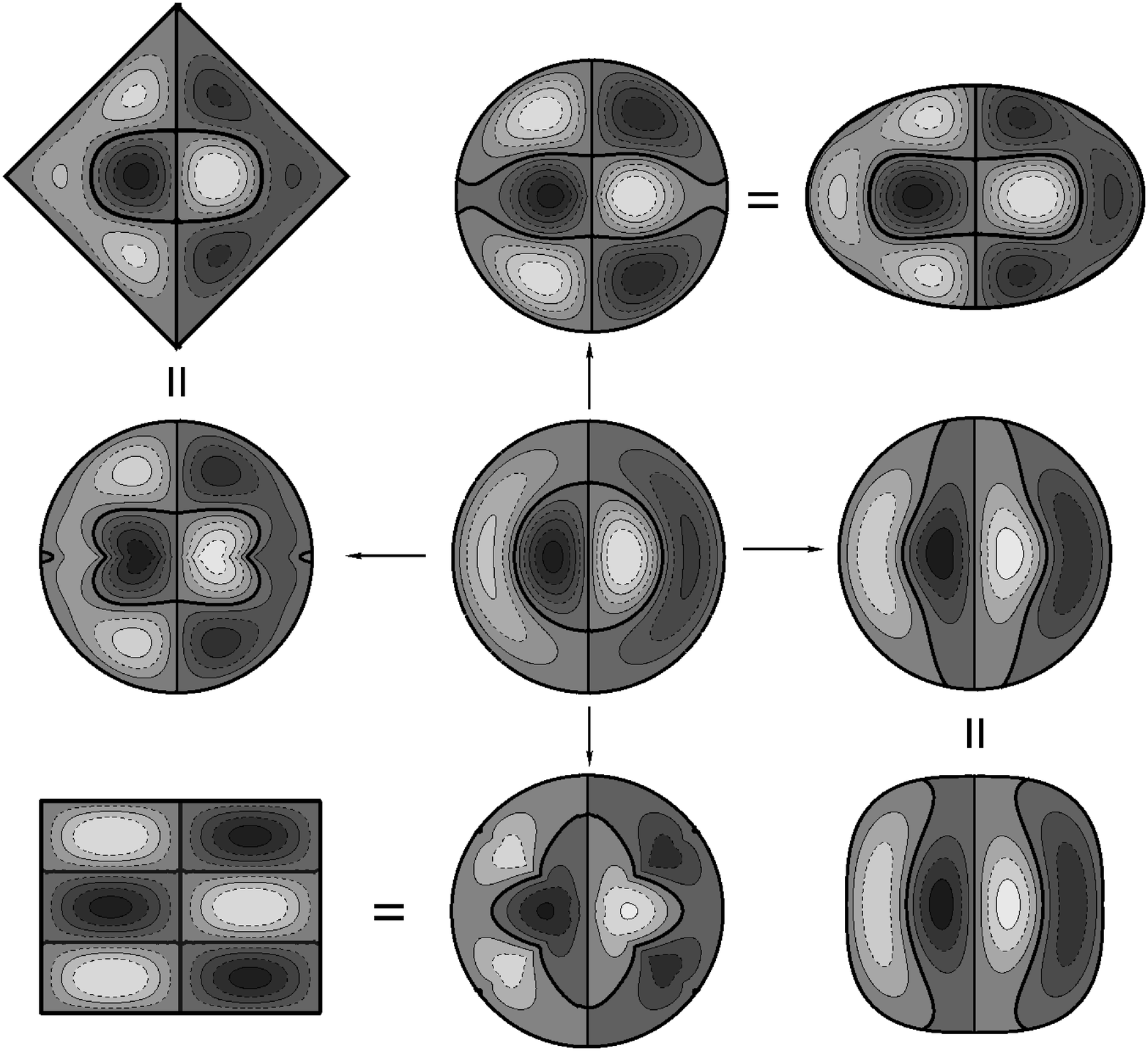}}}
\caption{Contours (dashed and continuous black lines) and nodal lines
  (thick solid black lines) of the wavefunction for a particular case,
  say, $l = 1$ and $ j = 2$. The figure in the center is that for a
  circular boundary of unit radius and is the starting point for all
  the four cases shown here. The figures pointed by the arrows in the
  right, left, up and down directions are the contours and nodal lines
  of the metric deformed wavefunction in a circular boundary and
correspond to supercircular ($t = 3$), square ($t = 1$), elliptical
($\epsilon = 0.75 $, $a = 1$) and rectangular (of length is to
  width ratio $4/3$) boundaries respectively. The figures
  next to these at corners are the plots of wavefunctions in their original shape.}
\label{fig:1}
\end{figure}

In the $l \neq 0$ case, the first order wavefunction correction is given by
\begin{align}
&\psi_{l,j}^{(1)} =\, a_{0} J_{0}(\rho) + \frac{\rho N_{l,j}
  C_{l}}{2}J^{\prime}_{l}(\rho)+\left\{a_{l}J_{l}(\rho) + \frac{\rho
    N_{l,j} J^{\prime}_{l}(\rho)}{2}\left(C_{2l} +
  \frac{E_{l,j}^{(1)}}{E_{l,j}^{(0)}}\right)\right\} \cos (l\alpha)
  \nonumber\\
&\qquad\qquad+\sum_{\substack{p=1\\ p\neq l}}^{\infty} \left\{a_{p}J_{p}(\rho)  + \frac{\rho}{2} N_{l,j} J^{\prime}_{l}(\rho)\left(C_{l+p} + C_{|l-p|} \right)\right\}  \cos (p\alpha).
\label{eq:23}
\end{align}
Here we have considered only the `cosine' form of $\psi_{l,j}^{(0)}$
(see Eq. (\ref{eq:21})) for the $l \neq 0$ case. The other solution
with the $\sin (l\alpha)$ term can be treated similarly. We have
estimated the first order energy corrections by imposing the
respective boundary conditions for $i =1$ given in
Eqs. (\ref{eq:bctot}).  $E_{l,j}^{(1)}$ is also verified by
substituting $\psi_{l,j}^{(0)}$ in Eq. ({\ref{eq:16b}}). We have
\begin{align}
E_{l,j}^{(1)} = & -E_{l,j}^{(0)}C_{2l}  ~, ~\qquad \qquad\qquad \qquad \mbox{(DBC)}~;\nonumber \\
E_{l,j}^{(1)} = &
-E_{l,j}^{(0)}C_{2l}\left(\frac{\rho^{\prime^{2}}_{_{l,j}} +
  l^{2}}{\rho^{\prime^{2}}_{_{l,j}} - l^{2}}\right) ~,~~~\qquad\mbox{(NBC)}~,\nonumber  
\end{align}
where the corresponding $E_{l,j}^{(0)}$ are given by
Eq. ({\ref{eq:endbc}}) and Eq. ({\ref{eq:ennbc}}) respectively. This
is generally non-vanishing unlike the earlier case. The second order
energy correction becomes crucial when $C_{2l} = 0$. The choice of the `sine' solution
for $\psi_{l,j}^{(0)}$ (Eq. {\ref{eq:21}}) gives
\begin{align}
E_{l,j}^{(1)} = &~ E_{l,j}^{(0)}C_{2l}  ~, \qquad \qquad\qquad \quad
\mbox{(DBC)}~; \nonumber \\
E_{l,j}^{(1)} = &~ E_{l,j}^{(0)}C_{2l}\left(\frac{\rho^{\prime^{2}}_{_{l,j}} +
  l^{2}}{\rho^{\prime^{2}}_{_{l,j}} - l^{2}}\right) ~,~~\quad\mbox{(NBC)}~.\nonumber  
\end{align}
Further, the coefficients $a_{0}$ and $a_{p}$ (for $p \neq 0,~l$) are obtained as a bonus
giving,
\[ \left.\hspace{-2.6cm}
\begin{array}{cl}
a_{0}= & -\frac{N_{l,j} \rho_{_{l,j}} C_{l}}{2}\frac{J^{\prime}_{l}(\rho_{_{l,j}})}{J_{0}(\rho_{_{l,j}})} \\
a_{p}= & -\frac{N_{l,j}
  \rho_{_{l,j}}J^{\prime}_{l}(\rho_{_{l,j}})}{2J_{p}(\rho_{_{l,j}})}\left[C_{p+l}
  + C_{|p-l|} \right]\phantom{(\rho^{\prime^{2}}_{l,j}-pl) C_{|p-l|}^{(1)}} \\
a_{l}= & 0 
\end{array}
\hspace{-2.6cm}\right\}\mathrm{(DBC)}\]
\[ \left.
\begin{array}{cl}
a_{0}= & -\frac{N_{l,j} \rho^{\prime}_{l,j} C_{l}}{2}\frac{J_{l}(\rho^{\prime}_{l,j})}{J_{1}(\rho^{\prime}_{l,j})}\\
a_{p}= & \frac{N_{l}J_{l}(\rho^{\prime}_{l,j})}{2\rho^{\prime}_{l,j}J_{p}^{\prime}(\rho^{\prime}_{l,j})}
\left[(\rho^{\prime^{2}}_{l,j}+pl)C_{p+l}^{(1)}
  +(\rho^{\prime^{2}}_{l,j}-pl) C_{|p-l|}^{(1)} \right]\\
a_{l}=& \frac{l^4}{(\rho^{\prime^{2}}_{l,j}-l^2)^2}C_{2l}
\end{array}
\right\}\mathrm{(NBC)}\] The coefficient $a_{l}$ is calculated from
the normalisation of the corrected wavefunction up to first order.
These results are consistent with the ones obtained in earlier
investigations \cite{b.6a,b.6b,b.6c}. The contours and the nodal lines
of a wavefunction corrected up to first order for different boundary
geometries are shown in Fig. \ref{fig:1}. The small change in the
nodal lines is visible for the case of supercircle deformation whereas
for the other cases, viz. square, rectangle and ellipse, the changes
are violent. At first glance it seems that one gets a wrong eigenmode
for the square in the upper left corner of the Fig. \ref{fig:1}. A
closer look shows that it is indeed an eigenfunction of a square
membrane where two degenerate modes, viz. (1,4) and (4,1), are mixed
in equal proportion with a relative negative sign. Under deformation
among the nodal lines only the line of symmetry is preserved as can be
seen from the examples of Fig. \ref{fig:1}. Moreover the number of
crossings of the nodal lines is also not conserved for such violent
perturbation. What seems to be preserved between the equivalent domains
is the number of humps and valleys.

\section{Results and Discussions} 
We next apply the analytical formalism developed in the earlier section to a few
specific boundary geometries. We have compared our perturbative
results against the numerical solutions obtained by using the Partial
Differential Equation Toolbox${\rm ^{TM}}$ of MATLAB$\TReg $. To
ensure the convergence of the eigenvalues in numerical method we have
restricted only to convex domains. We have considered the case of a
supercircle and an ellipse. The polar form for these two families of
curves are respectively given by 
\begin{align}
r(\theta)&=\frac{a}{(|\cos\theta|^{t}+|\sin\theta|^{t})^{1/t}}~,\label{eq:sc}\\
r(\theta)&=\frac{a \sqrt{1- \epsilon^2}}{\sqrt{1- \epsilon^2 \cos^2\theta}}~.\label{eq:elp}
\end{align}
The parameters defining the boundaries are $(a>0,t\geq1)$ and
$(a>0,\epsilon>0)$ respectively. Eq. (\ref{eq:sc}) defines a diamond
($45^{\circ}$ rotated square) for $t=1$, a circle for $t = 2$, a
supercircle for $t>1$ and $t \neq 2$ and a square as $t\rightarrow
\infty$. The specific form of $r(\theta)$ for these closed curves are
used to calculate the metric deformation and thereby estimate energy
corrections. These curves are chosen because they have a reflection
symmetry about y-axis and hence they can be represented by a Fourier
series given in Eq. \eqref{eq:19} with only cosine terms. The
perturbative prescription is seen to converge with dominant non-zero
corrections coming from the first few orders. It can be seen that the
first and second order corrections in energy are linear and bi-linear
in $C_{n}$ respectively. Further, it is clear that $E^{(m)}$ will be
$m$-linear in $C_{n}$, hence convergence of the Fourier coefficients,
$C_{n}$, will ensure the convergence of the series. In this analytic
formalism the approximation appears only through truncation of the
series given by Eqs. ({\ref{eq:12}}).

\begin{figure}[htb]
    \centerline{\scalebox{1.4}{\includegraphics{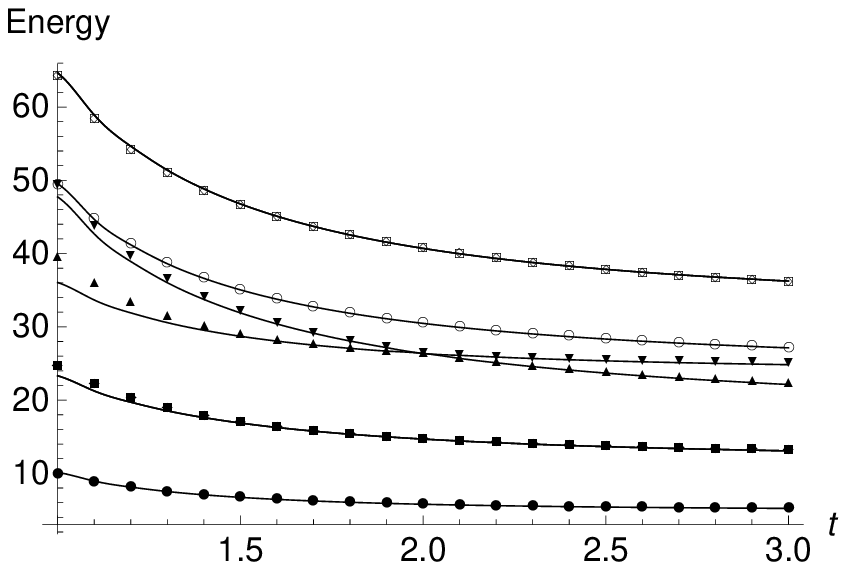}}}
    \caption{Comparison of the first few energy levels obtained
  numerically (denoted by points) and analytically (denoted by
  solid lines) for the supercircular boundary with the variation of the supercircular exponent ($t$) for fixed $a = 1$ for DBC.}
\label{fig:2}
\end{figure}

\begin{figure}[htb]
    \centerline{\scalebox{1.4}{\includegraphics{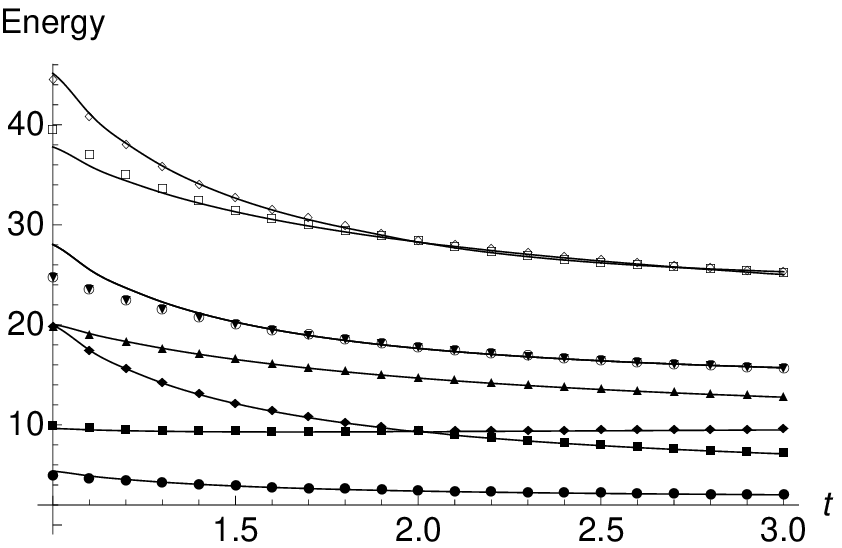}}}
    \caption{Comparison of the first few energy levels obtained
  numerically (denoted by points) and analytically (denoted by
  solid lines) for the supercircular boundary with the variation of the supercircular exponent ($t$) for fixed $a = 1$ for NBC.}
\label{fig:3}
\end{figure}

\begin{figure}[htb]
    \centerline{\scalebox{1.4}{\includegraphics{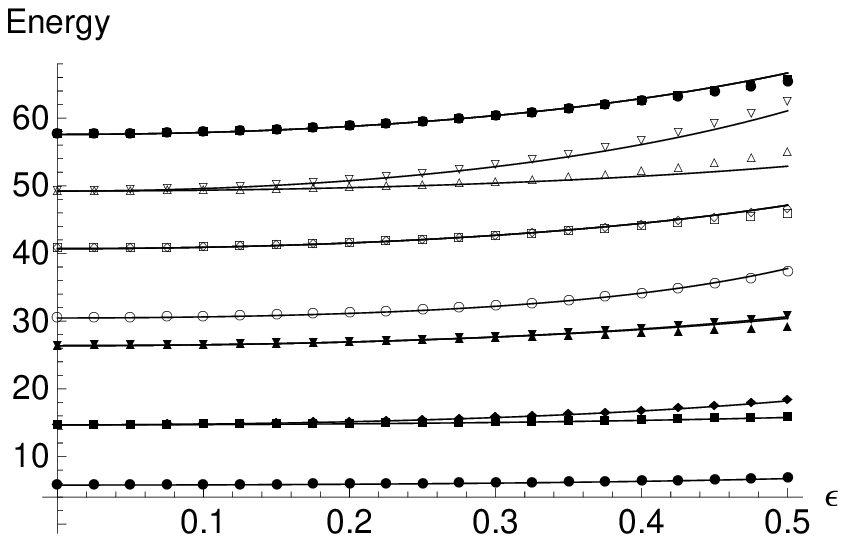}}}
    \caption{Comparison of the first few energy levels obtained
  numerically (denoted by points) and analytically (denoted by
  solid lines) for the elliptical boundary with the variation of the
  eccentricity ($\epsilon$) for fixed $a = 1$ for DBC.}
\label{fig:4}
\end{figure}

\begin{figure}[htb]
    \centerline{\scalebox{1.4}{\includegraphics{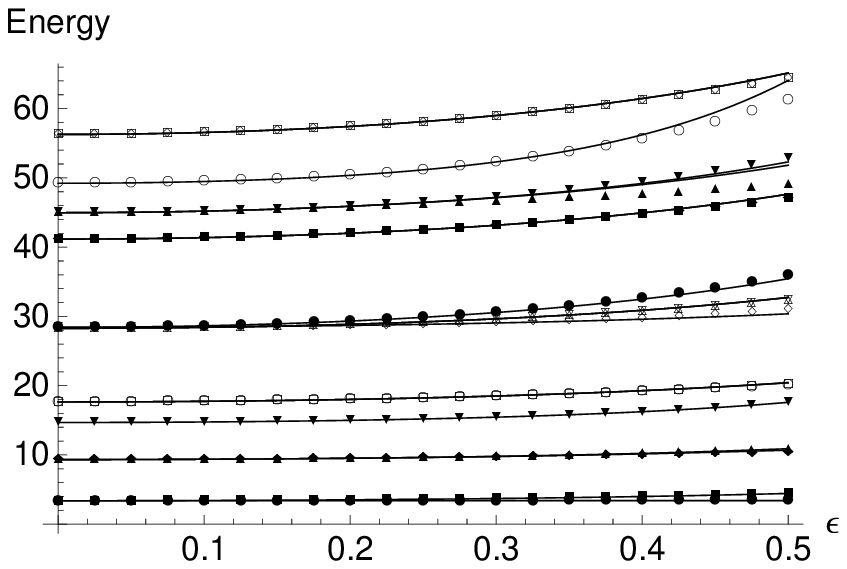}}}
    \caption{Comparison of the first few energy levels obtained
  numerically (denoted by points) and analytically (denoted by
  solid lines) for the elliptical boundary with the variation of the
  eccentricity ($\epsilon$) for fixed $a = 1$ for NBC.}
\label{fig:5}
\end{figure}

We have estimated energy corrections up to the second order in
perturbation for the $l = 0$ states and only first order corrections
for the $l \neq 0$ states. The two-fold degeneracy of the original $l
\neq 0$ states splits at the first order for $C_{2l} \neq 0$. In
Figs. \ref{fig:2} and \ref{fig:3} we have illustrated the comparison
of the analytical values calculated by our perturbation scheme with
their respective numerical ones for the supercircular boundary in the
range of $t$ from $1$ to $3$ for first few energy levels. Our results
are in good agreement with the numerical ones. The discrepancy is $
\sim 1\%$ for the supercircle within the range $1.5\leq t \leq 3$ and
is relatively larger ($\sim 5\%$) as it tends toward the diamond
shape, i.e., $t=1$. This larger discrepancy is anticipated because a
square is a violent departure from a circle and this large deformation
is against the inherent ingredient of the perturbative
method. Furthermore, energy levels corresponding to some $l$ values
exhibit level crossing phenomenon as reported earlier
\cite{b.6b,b.6c}. It is clear from the Figs. \ref{fig:2} and
\ref{fig:3} that the overall matching for the DBC is outstanding
except for few occasions in case of square where only the first order
correction (for $l \neq 0$ case) is included. However, inclusion of
higher order corrections will definitely improve the accuracy of our
method. In contrast, for the case of NBC most of the low-lying states
with non-zero $l$ values do not even have the first order
correction. So, in such cases the error is distinct for the
square. The method is expected to yield better results for smooth
boundaries without vertices. The results for the supercircle even at
the first order show better accuracy than their second order
counterparts for the square. Similarly, the comparison for elliptical
boundaries is shown in Figs. \ref{fig:4} and \ref{fig:5}. Here also,
the agreement is highly satisfactory for a wide range of $\epsilon$
and the discrepancy is $\sim 2\%$. A typical comparison between the
results obtained by numerical method ($Ns$) or the exact solution
($Es$) and our perturbative scheme ($Ps$) is shown in the Table
\ref{tab:1}.

In conclusion, we note that Fourier decomposition of the boundary
asymmetry makes the method completely general and it holds good for a
wide variety of boundaries and for general boundary conditions
also. The main advantage of this method over the others is that it is
boundary condition free and has general closed form solutions at every
order of perturbation. Next it maintains the same simple form of
boundary condition at every order of perturbation making the
application of the boundary condition easier. In principle, the higher
order corrections could also be calculated exactly but they are
algebraically more complicated and tedious to evaluate. Since our
solutions of the wavefunction are general (i.e. independent of
boundary condition), the other mixed boundary conditions such as,
Cauchy\cite{b.8} or Robin\cite{b.9}, could also be applied to obtain
the corresponding spectrum easily.

\section{Acknowledgements}
SP would like to acknowledge the Council of Scientific and Industrial
Research (CSIR), India for providing the financial support. The
authors would like to thank S. Bharadwaj, S. Kar, A. Dasgupta, S. Das
and Ganesh T. for useful discussions and help. The authors would like
to thank the referee for critical comments and suggestions for
improving the text.

%


\setlength{\unitlength}{1cm}
\setlength{\unitlength}{1cm}

\setlength{\unitlength}{1cm}
\setlength{\unitlength}{1cm}

\begin{table}
\caption{Comparison of the first few energy eigenvalues with the
  magnitude of $\%$ error (= $\rvert\frac{Ns-Ps}{Ns}\lvert \times
  100\%$ or $\rvert\frac{Es-Ps}{Es}\lvert \times 100\%$) for a
  supercircle (with $t=3,~a=1$), an ellipse (with $\epsilon =
  0.5,~a=1$) and a tilted square (with $t=1,~a=1$).}
\begin{center}
\begin{tabularx}{\textwidth}{XXXXXXXXX}\toprule
\multicolumn{3}{c}{{\bf Supercircle}}&\multicolumn{3}{c}{{\bf Ellipse}}&\multicolumn{3}{c}{{\bf Square}}\\\toprule
\multirow{2}{*}{~~~$Ns$} &\multirow{2}{*}{~~~$Ps$} & ~~$\%$ & \multirow{2}{*}{~~~$Ns$} & \multirow{2}{*}{~~~$Ps$} & ~~$\%$ & \multirow{2}{*}{~~~$Es$}  & \multirow{2}{*}{~~~$Ps$}  &~~ $\%$   \\
& & Error& & & Error & & & Error \\ \toprule 
\multicolumn{9}{c}{{\bf Dirichlet Boundary Condition}}\\ \midrule 
~ 5.219  &~ 5.217 &0.04 &~ 6.744 &~ 6.744 &0.00&~9.870 & 10.129&2.62 \\
13.193 & 13.076 &0.89& 15.893 & 15.776  &0.74& 24.674 & 23.317&5.49 \\
13.202 & 13.076 &0.95& 18.339 & 18.218 &0.66& 24.674 & 23.317&5.49 \\
22.372 & 22.141 &1.03& 29.080 & 30.416 &4.59& 39.478 & 36.047&8.69 \\
25.088 & 24.838 &1.00& 30.725 & 30.652 &0.24& 49.348 & 47.726&3.29 \\
27.157 & 27.138 &0.07& 37.191 & 37.777 &1.58& 49.348 & 49.533&0.37 \\
36.042 & 36.254 &0.59& 45.762 & 47.115 &2.96& 64.152 & 64.648&0.77 \\
36.046 & 36.254 &0.58& 46.570 & 47.136 &1.22& 64.152 & 64.648&0.77 \\
44.449 & 43.835 &1.38& 54.937 & 52.886 &3.73& 83.892 & 78.166&6.82 \\
44.490 & 43.835 &1.47& 62.334 & 61.073 &2.03& 83.892 & 78.166&6.82 \\
50.967 & 51.203 &0.46& 65.343 & 66.662 &2.02& 88.826 & 87.429&1.57 \\  \midrule 
\multicolumn{9}{c}{{\bf Neumann Boundary Condition}}\\ \midrule 
~ 2.975  &~ 3.019 &1.48&~ 3.426 &~ 3.407 &0.55&~ 4.935 &~ 5.384&9.09 \\
~ 2.978  &~ 3.019 &1.38&~ 4.462 &~ 4.442 &0.45&~ 4.935 &~ 5.384&9.09 \\
~ 7.136  &~ 7.115 &0.29& 10.387 & 10.695 &2.97&~ 9.870 &~ 9.648&2.25 \\
~ 9.524  &~ 9.500 &0.25& 10.846 & 10.904 &0.53& 19.739 & 19.981&1.23 \\
12.788 & 12.779 &0.07& 17.563 & 17.596 &0.19& 19.739 & 20.059&1.62 \\
15.513 & 15.719 &1.33& 20.067 & 20.419 &1.75& 24.674 & 28.031&13.61~ \\
15.529 & 15.719 &1.22& 20.206 & 20.448 &1.19& 24.674 & 28.031&13.61~ \\
25.100 & 25.039 &0.24& 31.018 & 30.370 &2.09& 39.478 & 37.785&4.29  \\
25.190 & 25.315 &0.50& 32.288 & 32.733 &1.38& 44.413 & 45.141&1.64 \\
25.232 & 25.315 &0.33& 32.322 & 32.736 &1.28& 44.413 & 45.141&1.64 \\
25.246 & 25.327 &0.32& 35.969 & 35.442 &1.47& 49.348 & 52.028&5.43 \\ \bottomrule
\end{tabularx}
\label{tab:1}
\end{center}
\end{table}

\end{document}